\documentclass{aa}

\usepackage{graphicx}
\usepackage{txfonts}
\bibpunct{(}{)}{;}{a}{}{,} % to follow the A&A style

\begin{document}

   \title{Bayesian radio interferometric imaging with direction-dependent calibration}

   \author{Jakob Roth
          \inst{1}\inst{2}\inst{3}
          \and
          Philipp Arras
		  \inst{1}\inst{3}
		  \and
		  Martin Reinecke
		  \inst{1}
        \and
        Richard A. Perley
        \inst{4}
        \and
        R\"udiger Westermann
        \inst{3}
		  \and
		  Torsten A. En{\ss}lin
		  \inst{1}\inst{2}
	  }

   \institute{Max Planck Institute for Astrophysics, Karl-Schwarzschild-Str. 1, 85748 Garching, Germany
         \and
             Ludwig-Maximilians-Universit\"at, Geschwister-Scholl-Platz 1, 80539 Munich, Germany
         \and
		 	Technische Universität M\"unchen (TUM), Boltzmannstr. 3, 85748 Garching, Germany
         \and
         National Radio Astronomy Observatory, P. O. Box 0, Socorro, NM 87801, USA
		 }

   \date{Received XXXX; accepted XXXX}

  \abstract
   {Radio interferometers measure frequency components of the sky brightness, modulated by the gains of the individual radio antennas. Due to atmospheric turbulence and variations in the operational conditions of the antennas these gains fluctuate. Thereby the gains do not only depend on time but also on the spatial direction on the sky. To recover high quality radio maps an accurate reconstruction of the direction and time-dependent individual antenna gains is required.}
   {This paper aims to improve the reconstruction of radio images, by introducing a novel joint imaging and calibration algorithm including direction-dependent antenna gains.}
   {Building on the \texttt{resolve} framework, we designed a Bayesian imaging and calibration algorithm utilizing the image domain gridding method for numerically efficient application of direction-dependent antenna gains. Furthermore by approximating the posterior probability distribution with variational inference, our algorithm can provide reliable uncertainty maps.}
   {We demonstrate the ability of the algorithm to recover high resolution high dynamic range radio maps from VLA data of the radio galaxy Cygnus~A. We compare the quality of the recovered images with previous work relying on classically calibrated data. Furthermore we compare with a compressed sensing algorithm also incorporating direction-dependent gains.}
   {Including direction-dependent effects in the calibration model significantly improves the dynamic range of the reconstructed images compared to reconstructions from classically calibrated data. Compared to the compressed sensing reconstruction, the resulting sky images have a higher resolution and show fewer artifacts. For utilizing the full potential of radio interferometric data, it is essential to consider the direction dependence of the antenna gains.}


   \maketitle

\section{Introduction}
Over the last decades, many modern radio interferometers like the Very Large Array (VLA), the Atacama Large Millimeter Array (ALMA), the Low-Frequency Array (LOFAR), the Australian Square Kilometer Array Pathfinder (ASKAP), and MeerKAT have been constructed and are now providing high-quality data.
With \mbox{LOFAR 2.0}, MeerKAT+, and the Square Kilometre Array (SKA), the future will bring even more instrumental progress to radio astronomy.
Utilizing the full potential of these instruments has become a major challenge, not only because of the enormous data size but also because of the complexity and variety of processes affecting the measurement. 

This makes the accurate calibration of the antenna gains an important aspect for unraveling the full information from the measurement data.
The antenna gains are complex-valued functions, describing the sensitivity of the individual antennas of the interferometer, and depend on time and spatial direction on the sky.
Traditionally direction-dependent effects (DDE) were neglected, and only direction-independent effects (DIE) were considered, rendering the antenna gains only time-dependent.

Clearly, DIEs are the dominant contribution to the antenna gains; nevertheless, DDEs exist, and neglecting them in the calibration procedure limits the image fidelity and dynamic range of the resulting sky reconstructions.
From the instruments mentioned above, especially LOFAR measurements are affected by DDEs in the antenna gains, making a DDE calibration essential.
However, also for other instruments correcting for DDEs is beneficial.

Historically, the first DDE considered in radio interferometry was the so called w-effect resulting from non-coplanar baselines of the interferometer.
Many algorithms have been developed to account for the w-effect (as for example \cite{Cornwell1992}, \cite{Perley1999}, \cite{Cornwell2008}, \cite{Offringa2014}, \cite{Arras2021_gr}).
Since the w-effect is purely geometrical it can be computed form the antenna configuration and, in contrast to DDEs arising form atmospheric fluctuations and operational conditions of the antennas, it does not need to be reconstructed from observational data.
For the rest of this paper we will focus on DDEs which are not known a priori but need to be recovered from data, and which might also be time-dependent.

Calibration algorithms correcting for such unknown and potentially time-dependent DDEs have been developed over the last years, initially building on w-effect correction techniques as for example proposed by \cite{Bhatnagar2008}. Nowadays DDE correction is most commonly achieved via partitioning the field of view into small facets, assuming piecewise constant DDEs.
Within each facet, a DIE calibration is performed.
This faceting based approach was developed in \cite{Smirnov_2015}, \cite{Weeren_2016}, and \cite{Tasse_2018}.
\cite{Albert2020} extended this approach by proposing to infer a smooth phase screen within each facet.

\cite{Repetti_2017} introduced a joint DDE calibration and imaging algorithm going beyond faceting and the assumption of piecewise constant DDEs.
For regularization, smoothness of the DDEs is enforced via modeling them as band limited Fourier kernels.
Additionally, the sky brightness is regularized via compressed sensing, assuming sparsity in an orthonormal basis.
This approach was generalized in \cite{Thouvenin2018}, \cite{Dabbech2019} and \cite{Birdi_2020}, adding the time dependence to the DDE correction and also accounting for polarization.
Building on this algorithm \cite{Dabbech_2021} demonstrated the benefit of DDE calibration on a Cygnus~A VLA dataset.

In \cite{Arras_2019} a joint Bayesian imaging and DIE calibration method was presented. Here we extend this approach to a joint Bayesian imaging and DIE as well as DDE calibration method. Similar to \cite{Repetti_2017} we a priori assume smoothness of the DDEs.
Nevertheless, we do not fix the degree of smoothness via a band limited kernel but use self-adaptive correlations priors described in Section \ref{sec:gains}. Also for the DIE gain solutions and the sky brightness we use self-adaptive correlation priors.
As in \cite{Dabbech_2021}, we also demonstrate the method presented in this paper on Cygnus~A VLA data.
This allows for insightful comparisons and validations of the two algorithms. 

For computational efficiency, we utilize the recently developed image domain gridding algorithm \citep{Tol_2018}. Image domain gridding allows for an efficient evaluation of the radio interferometric measurement equation in the presence of direction-dependent antenna gains. Therefore image domain gridding is a key component for correcting DDEs not having band limited kernels.

The remainder of the article is structured as follows. In section \ref{sec:InfProblem} we outline the radio imaging and calibration problem in a Bayesian setting. Section \ref{sec:algo} describes our reconstruction algorithm. In section \ref{sec:demo} we demonstrate the algorithm and compare the image quality to existing work. Section \ref{sec:conclusion} gives a conclusion and an outlook to future work.

\section{The Inference Problem}\label{sec:InfProblem}
The radio interferometric measurement equation, derived in \cite{Smirnov_2011}, relates a given sky brightness and antenna gains to model visibilities.
To be more precise, the model visibility $\tilde{V}_{pqt}$ of antennas $p$ and $q$ at time $t$ is given by
\begin{align}\label{eq:rime}
   \tilde{V}_{pqt} = \int C(\boldsymbol{l},w_{pqt})I(\boldsymbol{l})G_{p}(\boldsymbol{l},t)G^{*}_{q}(\boldsymbol{l},t) e^{-2\pi i (\boldsymbol{k}_{pqt}\cdot\boldsymbol{l})} d\boldsymbol{l},
\end{align}
with
\begin{itemize}
   \item $\boldsymbol{l} = (l,m)$ being the sky coordinates,
   \item $t$ the time coordinate,
   \item $\boldsymbol{k}_{pqt} = (u_{pqt}, v_{pqt})$ the baseline uv-coordinates in units of the imaging wavelength,
   \item $C(\boldsymbol{l},w_{pqt}) = \exp{\left(-2\pi i w_{pqt}\left(\sqrt{1-\boldsymbol{l}^2} - 1\right)\right)} / \sqrt{1-\boldsymbol{l}^2}$ the $w$-effect,
   \item $I(\boldsymbol{l})$ the sky brightness,
   \item $G_{p}(\boldsymbol{l},t)$ the gain of antenna $p$.
\end{itemize}
Thus, model visibilities are Fourier components of the sky brightness $I$ modulated by the antenna gains $G_p$ and $G_q$ as well as by $C$, encoding the w-effect. The visibilities $V_{pqt}$ actually measured by the interferometer are the noise less visibility $\tilde{V}_{pqt}$ plus some additive noise $n_{qpt}$
\begin{align}\label{eq:rspn}
   V_{pqt} = \tilde{V}_{pqt} + n_{qpt}.
\end{align}
While the w-effect $C$ is known, the sky brightness $I$ as well as the antenna gains $G$ have to be reconstructed.
Recovering the underlying sky brightness $I$ and antenna gains $G$ from the noisy measurement data is an inverse problem. 

In this work, we solve this inverse problem in a probabilistic setting. Thus, we seek for an estimate of the posterior distribution $P(I,G|V)$, which is the probability distribution of the sky $I$ and antenna gains $G$ conditioned on the measured visibilities $V$. Via Bayes' Theorem 
\begin{align}\label{eq:bayes}
   P(I,G|V) = \frac{P(V|I,G)P(I,G)}{P(V)},
\end{align}
this posterior distribution can be expressed in terms of the likelihood $P(V|I,G)$, the prior $P(I,G)$, and the evidence $P(V)$. While the likelihood and the prior actually depend on $I$ and $G$, the evidence only acts as a normalization constant. The prior $P(I,G)$ encodes general knowledge and assumptions about the sky brightness $I$ and the antenna gains $G$ before considering the data. Based on these assumptions, the prior assigns a probability density to each possible configuration of $I$ and $G$. We will discuss the prior assumptions of this algorithm in detail in section \ref{sec:prior}.

For a given $I$ and $G$ the model visibilities $\tilde{V}$ can be computed via the measurement equation \ref{eq:rime}. The likelihood $P(V|I,G) = P(V|\tilde{V})$ is determined by the statistics of the measurement noise in equation \ref{eq:rspn}. We will discuss the likelihood and the noise model in section \ref{sec:likelihood}.

\section{The Algorithm}\label{sec:algo}
\subsection{Prior Model}\label{sec:prior}
The sky brightness $I(\boldsymbol{l})$ is a largely continuous function of the spatial position $\boldsymbol{l}$ on the sky, except for locations of point sources. Similarly, the antenna gains $G_{pt}(\boldsymbol{l},t)$ are largely continuous functions of position $\boldsymbol{l}$ and time $t$.
Unless some parametric form is considered, continuous quantities have an infinite number of degrees of freedom.
Thus, reconstructing them from finite measurement data is an ill-posed problem, since the measurement data does not uniquely determine the solutions for $I(\boldsymbol{l})$ and $G_{pt}(\boldsymbol{l},t)$.
To circumvent this problem, some form of regularization is required.

Information field theory \citep{Ensslin_2019} gives the theoretical foundation for expanding Bayesian inference to continuous quantities. In Bayesian inference, classical regularizers do not exist. Instead, the prior allows to discriminate between all solutions compatible with the data. We make some mild assumptions about the involved quantities to derive a reasonable prior distribution. The key prior assumptions are smoothness and positivity, which we will detail in sections \ref{sec:cor} and \ref{sec:sky}, respectively.

We encode the prior probability distributions in the form of standardized generative models \citep{Knollmueller2018}. Standardized generative models transform a set of uniformly distributed Gaussian random variables $\{ \xi_i \}$ via a non-linear mapping to the desired target distribution.

\subsubsection{Smoothness}\label{sec:cor}
Almost all continuous physical quantities, called fields hereafter, are smooth functions of their respective coordinates. Thus, the values of a field at nearby locations are not independent of each other but are correlated. Exploiting such correlations enables the inference of fields from finite measurement data.

We model all correlations via Gaussian processes. A priori we assume statistical homogeneity and isotropy of the Gaussian process, meaning that a priori there is no special position or direction. According to the Wiener-Khintchine theorem \citep{Khintchine1934} the correlation kernels of homogeneous and isotropic processes are diagonal in Fourier space, with the power spectrum on the diagonal. We reconstruct the power spectrum of the field jointly with the field values themselves. For the power spectrum, we utilize a non-parametric model, encoding mild prior assumptions, i.e, it is falling and is itself a smooth function of the Fourier frequency. The details of the correlation model can be found in \cite{Arras2022}.

As we already pointed out in Section \ref{sec:prior}, we encode our prior distributions in the form of generative models. Also, this correlation model has the form of a generative model. Thus, it maps uncorrelated Gaussian random numbers $\xi$ to a correlated Gaussian process $\psi(\xi)$.

\subsubsection{Sky Prior Model}\label{sec:sky}
In the current version of the algorithm, we ignore polarization and only consider Stokes I imaging.
Since there is no negative flux, the sky brightness is constrained to be strictly positive $I(\boldsymbol{l}) > 0$.
In our model, $I(\boldsymbol{l}) = I_d(\boldsymbol{l}) + I_p(\boldsymbol{l})$ consists of two components, with $I_d(\boldsymbol{l})$ modeling the diffuse sky emission and $I_p(\boldsymbol{l})$ modeling the point source emission. Both have to obey the positivity constraint $I_{d/p}(\boldsymbol{l}) > 0$ individually.

The characterizing property of diffuse sky emission is that it is smoothly varying as a function of sky direction.
The brightness of the diffuse sky emission varies over several orders of magnitude and is strictly positive. For these reasons, we model the logarithm of the diffuse sky emission with a two-dimensional generative model for correlated Gaussian process $\psi_{sky}(\xi_{sky})$.
After exponentiation we get the generative model for $I(\boldsymbol{l})$
\begin{align}
   I_d = e^{\psi_{sky}(\xi_{sky})},
\end{align}
encoding a correlated log-normal distribution.
This satisfies the requirements of a strict positivity and smoothness. 

Also, the point source model should satisfy the positivity constraint. Furthermore, the point source model needs to be flexible enough to capture extreme variations in the brightness of point sources. Therefore we choose to model point sources with pixel-wise inverse gamma-distributed brightness values, as already done in \cite{Arras2021} and \cite{Selig2015}. We insert two such pixels with inverse gamma-distributed point source fluxes at the locations of the two sources near the core of Cygnus~A.

\subsubsection{Antenna Gain Prior Model}\label{sec:gains}
At the current stage of the algorithm, we neglect polarization leakage. Therefore the gain of antenna $p$ can be described as a diagonal matrix
\begin{align}
   G_{p}(\boldsymbol{l}, t) = 
   \begin{pmatrix}
      G^{rr}_{p}(\boldsymbol{l},t) & 0\\
      0 & G^{ll}_{p}(\boldsymbol{l},t)
   \end{pmatrix},
\end{align} 
with $G^{rr}_{p}(\boldsymbol{l},t)$ and $G^{ll}_{p}(\boldsymbol{l},t)$ being the calibrations of the two antenna feeds $\left<ll\right>$ and $\left<rr\right>$. The physical origin of the $\boldsymbol{l}$, t and jointly $(\boldsymbol{l},t)$-dependent variations of the antenna gains can be different.
For example, the purely time-dependent variations arise from fluctuations in the amplification of the antenna signals, as well as from atmospheric effects with angular scales larger than the antenna beam.
Only direction-dependent variation might arise from deviations between the actual antenna power pattern and the assumed model. Finally, pointing errors of the antennas and atmospheric effects with angular scales smaller than the antenna beam cause joint time and direction-dependent gain variations.

Thus, the physical origins of the time and/or direction dependent variations are different, and therefore, the statistics of the gain variations along these axes are expected to be different. For this reasons, we separate the purely time and direction-dependent terms from the joint direction and time-dependent term. Thus, we split $G^{ll/rr}_{pt}(\boldsymbol{l},t)$ into
\begin{align}\label{eq:gain_split}
   G^{ll/rr}_{p}(\boldsymbol{l}, t) = g^{ll/rr}_{p}(t) \cdot g^{ll/rr}_{p}(\boldsymbol{l}) \cdot g^{ll/rr}_{p}(\boldsymbol{l}, t),
\end{align}
with $ll/rr$ standing for the two cases $<ll>$ or $<rr>$.
Splitting the antenna gain into these three factors has the advantage that it allows to encode different prior knowledge into the generative models of the individual terms. For example, we can encode that the a priori expected fluctuations of $g^{ll/rr}_{p}(\boldsymbol{l})$ and $g^{ll/rr}_{p}(\boldsymbol{l}, t)$ are much smaller than that for $g^{ll/rr}_{p}(t)$.

In order to avoid degeneracies between the three antenna gain factors in  Eq. \ref{eq:gain_split}, we suppress the only time $t$ and only direction $\boldsymbol{l}$ dependent contributions in the generative model for $g^{ll/rr}_{p}(\boldsymbol{l}, t)$ by following a similar idea presented in \cite{Guardiani2022} which we will outline below.

Each antenna gain term is a complex-valued function.
Because the variations of phase and amplitude of the antenna gains have different origins, the typical temporal and spatial correlation length is expected to be different.
This makes it a natural choice to have distinct models for phase and amplitudes.
We model the logarithm of the gain amplitude and the phase with two distinct Gaussian processes.
Thus, we have the following forward model 
\begin{align}\label{eq:gain_mod}
   g^{ll/rr}_{p}(\cdot) = e^{\psi^{ll/rr}_{p\text{,logamp}}(\cdot) + i\psi^{ll/rr}_{p\text{,phase}}(\cdot)},
\end{align}
with $(\cdot)$ representing either $(t)$ or $(\boldsymbol{l})$.

As already mentioned we suppress the contribution of only time $t$ or only direction $\boldsymbol{l}$ dependent parts in the generative model for $g^{ll/rr}_{p}(\boldsymbol{l}, t)$ as they should be captured by $g^{ll/rr}_{p}(t)$ and $g^{ll/rr}_{p}(\boldsymbol{l})$ respectively.
To do so, we alternate the phase and amplitude models in Eq. \ref{eq:gain_mod} for the case of the joint $(\boldsymbol{l}, t)$-dependence.
Instead of directly using the output of the Gaussian processes $\psi^{ll/rr}_{\text{logamp}}(\boldsymbol{l}, t)$ and $\psi^{ll/rr}_{\text{phase}}(\boldsymbol{l},t)$ as a model for the jointly $(\boldsymbol{l}, t)$-dependent amplitudes and phases, we first subtract the integrals over $\boldsymbol{l}$ and $t$. Thus, we use $\tilde{\psi}^{ll/rr}_{(\cdot)}(\boldsymbol{l}, t)$ defined as
\begin{align}\label{eq:psi_prime}
   \tilde{\psi}^{ll/rr}_{(\cdot)}(\boldsymbol{l}, t) = \psi^{ll/rr}_{(\cdot)}(\boldsymbol{l}, t) &- (\text{vol}_{\boldsymbol{l}})^{-1} \cdot \int \psi^{ll/rr}_{(\cdot)}(\boldsymbol{l}, t) \, d\boldsymbol{l} \notag\\
   &- (\text{vol}_{t})^{-1} \cdot \int \psi^{ll/rr}_{(\cdot)}(\boldsymbol{l}, t) \, dt,
\end{align}
for the amplitude and phase models, with $(\cdot)$ standing for $(\text{logamp})$ or $(\text{phase})$ respectively. Thereby $\text{vol}_{\boldsymbol{l}} = \int 1\, d\boldsymbol{l}$ and $\text{vol}_{t} = \int 1 \, dt$ are the volumes of the direction and time spaces. Thus, the jointly $(\boldsymbol{l}, t)$-dependent calibration factor can be summarized as 
\begin{align}\label{eq:gain_mod_modified}
   g^{ll/rr}_{p}(\boldsymbol{l}, t) = e^{\tilde{\psi}^{ll/rr}_{\text{logamp}}(\boldsymbol{l}, t) + i\tilde{\psi}^{ll/rr}_{\text{phase}}(\boldsymbol{l}, t)},
\end{align}
with $\tilde{\psi}^{ll/rr}_{(\cdot)}(\boldsymbol{l}, t)$ as defined in Eq. \ref{eq:psi_prime}.

This definition of $\tilde{\psi}^{ll/rr}_{(\cdot)}(\boldsymbol{l}, t)$ reduces the contributions of only $t$ or only $\boldsymbol{l}$ dependent parts. To explicitly verify this, we can, as a test, add an only $\boldsymbol{l}$-dependent contribution ${\psi_{(\cdot)}^{'ll/rr}}(\boldsymbol{l}, t) = \psi_{(\cdot)}^{ll/rr}(\boldsymbol{l}, t) + \phi(\boldsymbol{l})$. Then, the resulting $\tilde{\psi}_{(\cdot)}^{'ll/rr}(\boldsymbol{l}, t)$ of the modified $\psi_{(\cdot)}^{'ll/rr}(\boldsymbol{l}, t)$ is:
\begin{align}
   \tilde{\psi}_{(\cdot)}^{'ll/rr}(\boldsymbol{l}, t) = \psi_{(\cdot)}^{ll/rr}(\boldsymbol{l}, t) + \phi(\boldsymbol{l}) &- (\text{vol}_{\boldsymbol{l}})^{-1} \cdot \int \left( \psi_{(\cdot)}^{ll/rr}(\boldsymbol{l}, t) + \phi(\boldsymbol{l}) \right) \, d\boldsymbol{l}  \notag\\
   &- (\text{vol}_{t})^{-1} \cdot \int \left( \psi_{(\cdot)}^{ll/rr}(\boldsymbol{l}, t) + \phi(\boldsymbol{l}) \right)  \, dt  \notag\\
   &=  \tilde{\psi}_{(\cdot)}^{ll/rr}(\boldsymbol{l}, t) + c.
\end{align}
Thus, only $\boldsymbol{l}$-dependent contributions, and similarly also only $t$-dependent contributions, in $\psi_{(\cdot)}^{ll/rr}(\boldsymbol{l}, t)$ imprint on $\tilde{\psi}_{(\cdot)}^{ll/rr}(\boldsymbol{l}, t)$ only as a constant offset $c$. This guides the reconstruction to capture only $t$ or only $\boldsymbol{l}$-dependent variations of the gain factor by $g^{ll/rr}_{p}(t)$ and $g^{ll/rr}_{p}(\boldsymbol{l})$, respectively.

\subsection{Forward Model}\label{sec:model}
Besides the science target, also calibration objects are observed. We combine the science observation with the calibration observations into a joint forward mode as described in the next three subsections.

\subsubsection{Science Target} For the science target observation, we use the antenna gain and sky models as described above and compute the corresponding model visibilities via the measurement equation \ref{eq:rime}. For the numeric evaluation of the measurement equation we use the image domain gridding algorithm \citep{Tol_2018} as outlined in Sec. \ref{sec:idg}.

\subsubsection{Flux Calibrator} For the flux calibration observation, we do not reconstruct the sky brightness but assume it to be a point source with known flux in the center of the field of view. Thereby, we assume the fluxes of the calibration sources reported in \cite{Perley2013}.

We use only the direction-independent factors in the antenna gain model, since direction-dependent calibration cannot be performed with a single point source. Furthermore, the flux calibration antenna gains only share the amplitudes with the other observations while having a separate phase model. Thus, the flux calibration antenna gains share the Gaussian process for the logarithm of the amplitude $\psi_{\text{logamp}}(t)$ but not the phase process $\psi_{\text{phase}}(t)$.
   
Additionally, we allow the flux calibration antenna amplitudes to have a small offset to the amplitudes of the science target. We assume this offset to be a piecewise constant function of time $t$ with the breaking points at the time steps where the telescope switches the target. To summarize, the antenna gain for the flux calibration is given by
\begin{align}
   {g^{ll/rr}_{p}(t)}_{\text{Fcal}} = e^{\psi_{\text{logamp}}^{ll/rr}(t) + \alpha_{\text{Fcal}}^{ll/rr}(t) + i{\psi_{\text{phase}}^{ll/rr}}_{Fcal}(t)}.
\end{align}
Here $\psi_{\text{logamp}}^{ll/rr}(t)$ is the only time-dependent Gaussian process modeling the logarithm of the amplitude. This part of the gain model is shared with the other observations. $\alpha_{Fcal}^{ll/rr}(t)$ is the piecewise constant amplitude offset, and ${\psi_{\text{phase}}^{ll/rr}}_{Fcal}(t)$ is the phase of the amplitude calibration gain.
We use a zero-centered Gaussian as a prior model for the amplitude offsets $\alpha_{Fcal}^{ll/rr}(t)$. We do not assume a fixed variance of this Gaussian but simultaneously reconstruct it from the calibration data. Thereby, we use an inverse gamma prior on variance. Via the heavy tails of the inverse gamma distribution, we encode the prior knowledge that the offset is probably small but can sometimes significantly deviate from zero.

\subsubsection{Phase Calibrator} Similar to the flux calibration observation, we assume the phase calibrator to be a point source in the center of the field of view. Nevertheless, we do not set a fixed brightness of this point source but reconstruct it from the data.

Again we only use the direction-independent factors of the antenna gain model, since direction-dependent calibration can not be done with a single point source. In contrast to the flux calibration, for the phase calibration not only the amplitude model but also the phase model of the antenna gains is shared with the other observations.

As already introduced for the flux calibration model, we allow for small discontinuities in the calibration solution at the timesteps where the telescope switches the observation target. Since for the phase calibration observation, not only the amplitude model but also the phase model is shared with the science target, we also allow for an offset in phases between the two observations. Expressed as a formula, the antenna gain model for the phase calibration is defined by:
\begin{align}
   {g^{ll/rr}_{p}(t)}_{\text{Phcal}} = e^{\psi_{\text{logamp}}^{ll/rr}(t) + \alpha_{\text{Phcal}}^{ll/rr}(t) + i{\psi_{\text{phase}}^{ll/rr}}(t) + i{\beta_{\text{Phcal}}^{ll/rr}(t)}}.
\end{align}
Here, $\psi_{\text{logamp}}(t)$ and $\psi_{\text{phase}}(t)$ are the Gaussian processes modeling the phase and the logarithm of the antenna gains, which are shared with the other observations. $\alpha_{\text{Phcal}}^{ll/rr}(t)$ and $\beta_{\text{Phcal}}^{ll/rr}(t)$ are the piecewise constant offset functions. As a prior for the offsets $\alpha_{\text{Phcal}}^{ll/rr}(t)$ and $\beta_{\text{Phcal}}^{ll/rr}(t)$, we again use a zero-centered Gaussian with an inverse gamma distributed variance, as already described for the flux calibrator offset function.

\subsubsection{Image domain gridding}\label{sec:idg}
The radio response $R(I, G)$ maps a given sky brightness and antenna gains to model visibilities as defined by the measurement equation \ref{eq:rime}. Essentially the radio response $R$ is a non-uniform Fourier transformation of the sky brightness modulated by the w-effect and the antenna gains. These modulations by the antenna gains are different for every baseline and timestep. They make classical algorithms for computing the radio response numerically inefficient if one includes the direction-dependence of the gains. The novel image domain gridding algorithm by \cite{Tol_2018} overcomes this challenge and enables a fast evaluation of $R$ even in the presence of direction-dependent gains. We use image domain gridding to compute the radio response $R$.

For a fast optimization of the joint posterior estimate of $I$ and $G$ (see Sec. \ref{sec:geoVI}), we not only need to evaluate $R(I, G)$ but also its derivative. More specifically, we implemented the image domain gridding in such a way that we can also apply the Jacobian of $R$ at the position $(I_0, G0)$ to perturbations of the sky $dI$ and the gains $dG$. Thus, we can compute:
\begin{align}
   DR_{I_0, G_0}(dI, dG) = \begin{pmatrix} \left. \frac{\partial R}{\partial I} \right|_{I_0, G_0} & \left. \frac{\partial R}{\partial G} \right|_{I_0, G_0} \end{pmatrix} \begin{pmatrix}
      dI \\
      dG
   \end{pmatrix} .
\end{align}
Furthermore, we also need to apply the adjoint Jacobian to perturbed visibilities $dV$. Thus, we can evaluate:
\begin{align}
   \left(DR_{I_0, G_0}\right)^{\dagger}  (dV) = \begin{pmatrix} \left. \frac{\partial R}{\partial I} \right|_{I_0, G_0}^{*} \\
      \left. \frac{\partial R}{\partial G} \right|_{I_0, G_0}^{*} \end{pmatrix} \begin{pmatrix}
      dV
   \end{pmatrix} .
\end{align}
The Jacobian and its adjoint are implicitly evaluated, thus they are neither stored nor applied as a full matrix.

\subsection{Likelihood Model}\label{sec:likelihood}
With the above forward model and the image domain gridding algorithm for the measurement equation \ref{eq:rime} the model visibilities can be computed.
The statistics of the noise in Eq. \ref{eq:rspn} then determines the likelihood.
We assume Gaussian measurement noise statistics.
Nevertheless we do not assume a fixed noise covariance, but allow for an outliers correction by introducing a noise correction factor per data point.
During the inference we jointly reconstruct these correction factors with the calibration and sky brightness. For each correction factor we use an inverse gamma distribution as a prior, having a strong peak at $1$. Thereby, we can ensure that if the data is preferentially explained by the model, and only if this is not possible is classified as an outlier. A similar procedure was introduced in \cite{Arras2021} and was named ``automatic weighting''.

For all targets we compute the likelihood $P(V|I, G)$ as described above. The overall likelihood is then the product of the individual likelihoods for calibration and science targets:
\begin{align}
   P({V}|I,G) = \prod_i P(V_i|I,G).
\end{align}

Because we include temporal smoothness into our model (see Section \ref{sec:gains}) the likelihoods contribute information on the calibration solutions not only for the exact time intervals of their observation, but also for other observations.
Thereby, the temporal correlation kernels, reconstructed for the gain factors, determine the extrapolation of the gain solution of the calibration observations to the science observation (and vice versa). Therefore, jointly inferring the calibration solution and the science image is beneficial.

\subsection{Posterior}\label{sec:geoVI}
In the sections above, we have developed a likelihood model $P(V|I, G)$ and encoded the prior distribution $P(I, G)$ via the generative form of $I$ and $G$. To actually obtain an image $I$ and antenna gains $G$ from the observed visibilities $V$, it is necessary to compute, or approximate, the posterior distribution $P(I, G|V)$. Directly computing the posterior distribution via Eq. \ref{eq:bayes} is impossible since the model involves too many parameters and is highly non-linear. Instead, we approximate the posterior distribution via a two-stage algorithm.

In the first stage, we compute the maximum a posteriori solution (MAP) for sky $I$ and the antenna gains $G$ by maximizing the joint probability $P(V, I, G) = P(V|I,G) \cdot P(I,G)$ of visibilities, sky brightness, and antenna gains. This gives us  already a decent reconstruction of the sky brightness and the calibration solutions; nevertheless, it lacks a reliable uncertainty quantification.

Therefore, in the second stage, we compute a more accurate approximation of the posterior probability distribution of the sky brightness using the antenna gain solutions obtained in the first stage of the inference. Thus, expressed as a formula, we approximate $P(I|G,V)$. We compute this posterior approximation using geometric variational inference (geoVI), a technique developed by \cite{Frank2021}. geoVI is an accurate variational inference technique capable of capturing non-Gaussian posterior statistics and intercorrelation between the posterior parameters. The output of the geoVI inference is a set of posterior samples of the sky brightness. The mean sky brightness and its standard deviation can be computed from these samples.

Of course, ideally, not only the sky $I$ but also the antenna gains $G$ should be inferred via geoVI to include calibration uncertainties into the uncertainty quantification of the recovered sky. However, the computational demand of the geoVI inference is significantly higher than computing the MAP solution. This renders it at the current stage infeasible to have a joint geoVI estimate of the sky and the antenna gains. Therefore, in this work, the uncertainties on the sky brightness are only relative to the antenna gain calibration computed by MAP. Nevertheless, we believe these uncertainty maps to be significantly more accurate than noise estimates obtained from residual flux images of \texttt{clean}-like algorithms, also because residual images are always conditional to the assumed calibration solutions.

\section{Demonstration on Cygnus~A VLA Data}\label{sec:demo}
We demonstrate the performance of our algorithm on a VLA Cygnus~A observation. We use the S band data at $2.052\,\text{GHz}$ in all four VLA configurations. In recent years the Cygnus~A VLA data has been widely used for presenting and benchmarking novel imaging and calibration algorithm (see, for example, \citealt{Arras2021}, \citealt{Dabbech_2021}, \citealt{Sebokolodi_2020}, \citealt{Dabbech2018}). This allows for insightful comparison and validation of the imaging algorithms. In particular, we will compare the results of this work with the \texttt{resolve} reconstruction from \cite{Arras2021} and the compressed sensing reconstruction from \cite{Dabbech_2021}. In the following two subsections, we will briefly summarize these two algorithms. The code underlying the shown reconstructions can be found here: \url{https://gitlab.mpcdf.mpg.de/ift/public/cygnus_a_dde_cal}. The results of the sky reconstruction are available in fits format at \url{https://doi.org/10.5281/zenodo.7885385}.

\subsection{\texttt{resolve} with traditional calibration}
The \texttt{resolve} reconstruction of \cite{Arras2021} uses the same underlying framework and sky model as this work. Nevertheless, in contrast to this work, it uses traditionally calibrated data as an input, ignoring direction-dependent effects. This comparison allows studying the impact of direction-dependent calibration on Bayesian imaging.

In \cite{Arras2021}, the old \texttt{resolve} reconstruction without direction-dependent calibration was compared to single and multi-scale \texttt{clean} images, showing that the images reconstructed with \texttt{resolve} had a higher resolution than the single and multi-scale \texttt{clean} images. By comparing with higher frequency data, the small scale emission additionally recovered by the \texttt{resolve} algorithm could be confirmed to be real.

\subsection{Compressed sensing with direction-dependent calibration}
Furthermore, we compare with the algorithm of \cite{Dabbech_2021}. Similar to our method, \cite{Dabbech_2021} also performs a direction-dependent calibration avoiding faceting. Nevertheless, the methodology of this method is quite different as it uses compressed sensing for the regularization of the sky brightness and band limited Fourier kernels for the direction-dependent gains. Since the methodology of this method is very different is it well suited as a confirmation of the results of this work.  

\subsection{Comparison of Cygnus~A images}
\begin{figure*}
   \centering
      \includegraphics[width=17cm]{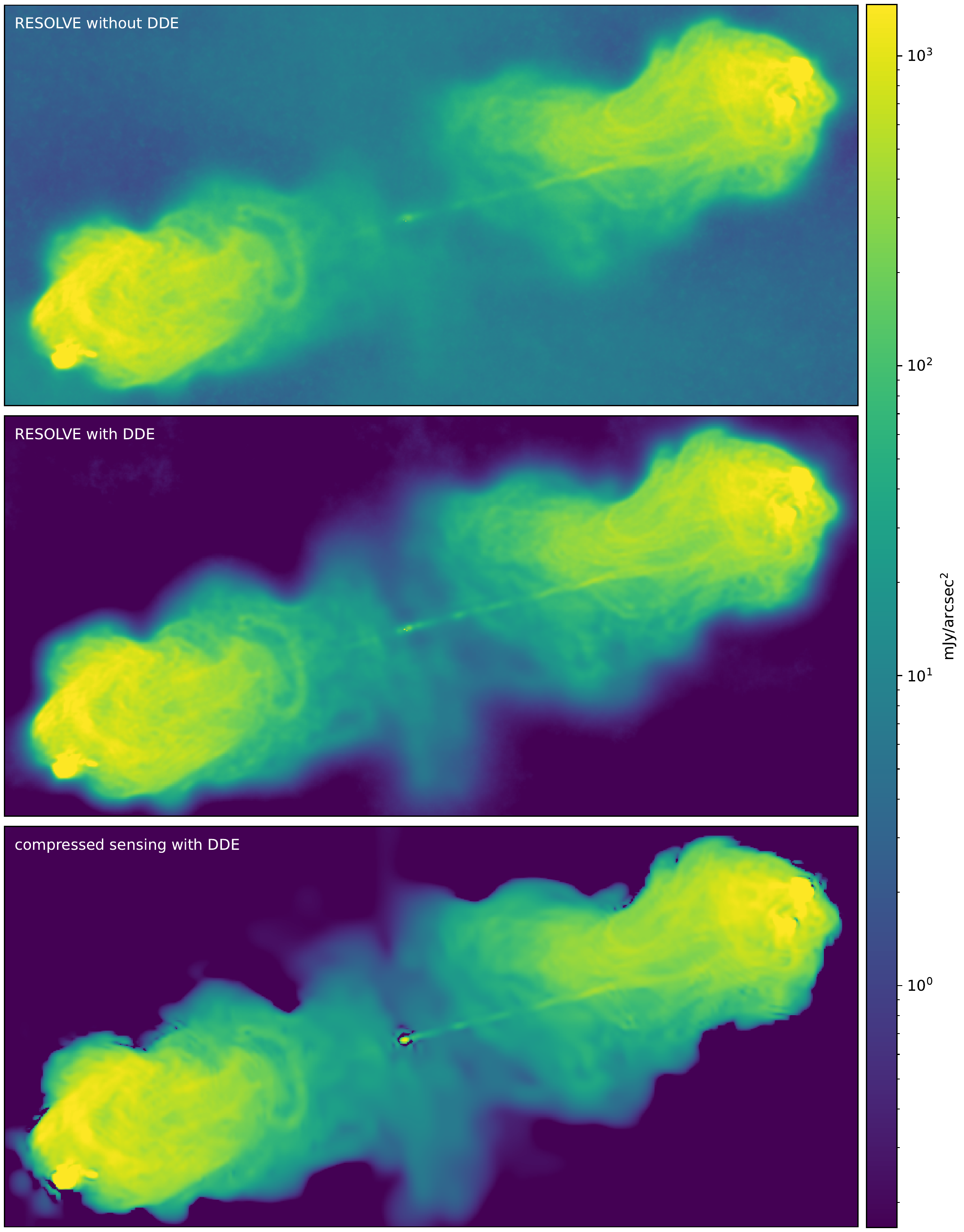}
        \caption{Comparison of Cygnus~A images with three algorithms. Top panel: Cygnus~A image from \cite{Arras2021} obtained with \texttt{resolve} without direction-dependent calibration. Central panel: The Cygnus~A image reconstructed with the algorithm of this work. Bottom panel: Cygnus~A image from \cite{Dabbech_2021} recovered by a compressed sensing algorithm including direction-dependent calibration. The colorbar is clipped for better visibility.}
        \label{fig:cyg_comp}
\end{figure*}

In Fig. \ref{fig:cyg_comp}, the Cygnus~A image reconstructions of this algorithm and the two studies mentioned above are displayed. All three algorithms deliver similar images of Cygnus~A. Especially the two Bayesian \texttt{resolve} reconstructions have only minimal artifacts in the recovered images. The compressed sensing algorithm produces some artifacts at the hot spots and the core of Cygnus~A.

An important quantity for comparing the quality of radio maps is their resolution, thus the smallest length scale at which features can still be imaged. While for classical \texttt{CLEAN} images, this length scale is usually set by the size of the \texttt{CLEAN} bean, more advanced imaging algorithms such as \texttt{resolve} and compressed sensing techniques can achieve super-resolution (\citealt{Arras2021}, \citealt{Dabbech2018}). Thus, for sufficiently high signal-to-noise, these algorithms can recover features smaller than the intrinsic resolution of the interferometer.

\begin{figure*}
   \centering
      \includegraphics[width=17cm]{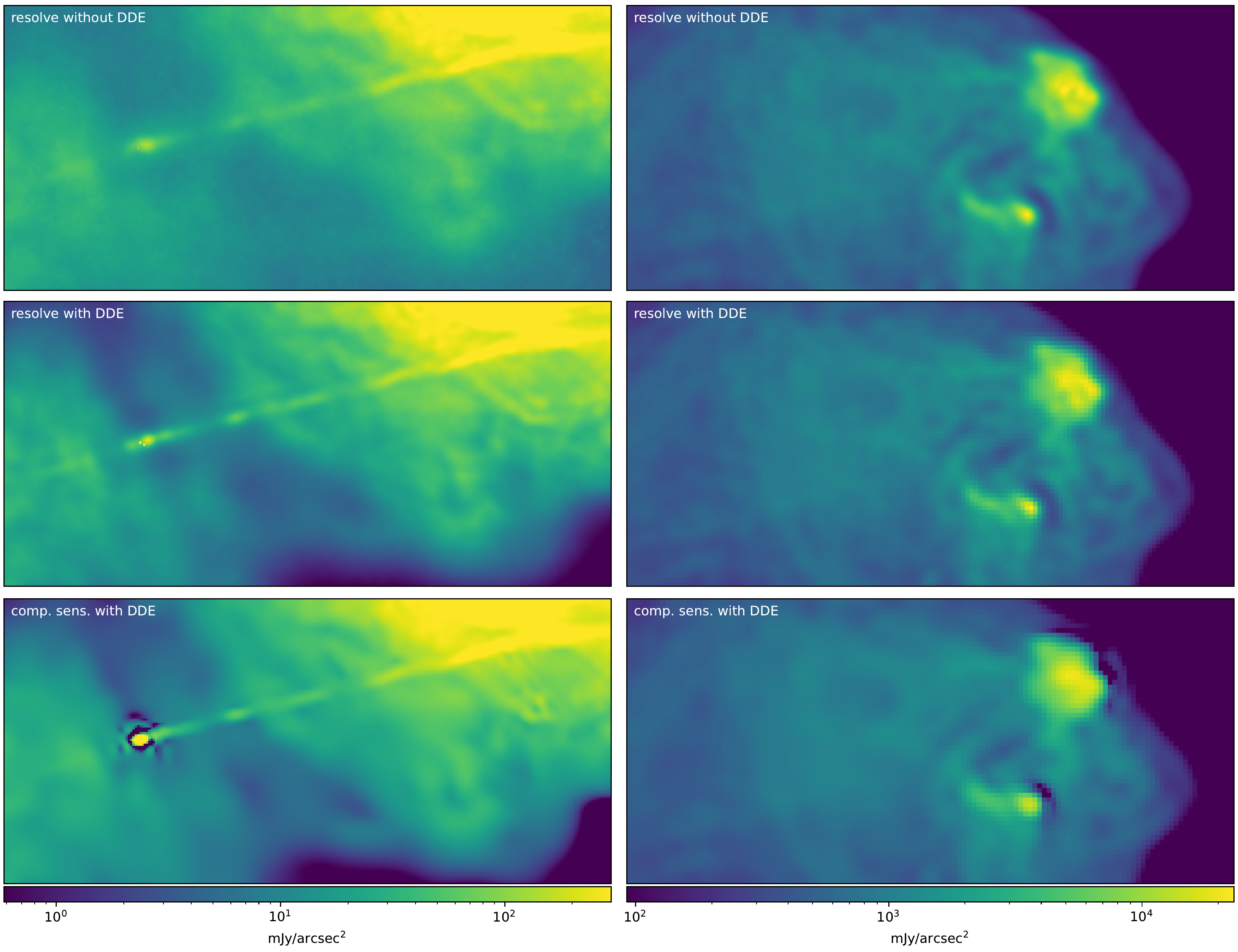}
        \caption{Left column: Zoom on the core and jet. Top panel: \cite{Arras2021}; Central panel: This work; Bottom panel: \cite{Dabbech_2021}. Right column: Zoom on the hot spots of western lobe. Top panel: \cite{Arras2021}; Central panel: This work; Bottom panel: \cite{Dabbech_2021}. The colorbars of the left and right columns are clipped and adapted to the respective surface brightness range.}
        \label{fig:zoom}
\end{figure*}

In high surface brightness regions, the resolution of both Bayesian images is slightly higher. Already in \cite{Arras2021}, the very high resolution of the Bayesian reconstruction compared with \texttt{clean} was discussed, and its fidelity was confirmed by comparison with higher frequency data. In order to depict the difference in resolution more clearly, the right column of Fig. \ref{fig:zoom} shows a zoom-in on the western hot spot. Also, in the zoom on the hot spots, the three algorithms yield very compatible reconstructions.

Fig. \ref{fig:zoom} left column depicts the core, the jet, and regions of the lobes with lower surface brightness. In low surface brightness regions, the resolution of the \texttt{resolve} reconstruction from \cite{Arras2021} using classically calibrated data is significantly lower than the resolution of the reconstruction of this work and also lower than the resolution of the compressed sensing reconstruction of \cite{Dabbech_2021}. Both algorithms utilizing direction-dependent calibration depict the sharp structures of the jet much more clearly than the algorithm of \cite{Arras2021} building on classical calibration. That direction-dependent calibration enhances the jet was already found in \cite{Dabbech_2021} compared to other methods building on classical calibration.

\begin{figure*}
   \centering
      \includegraphics[width=17cm]{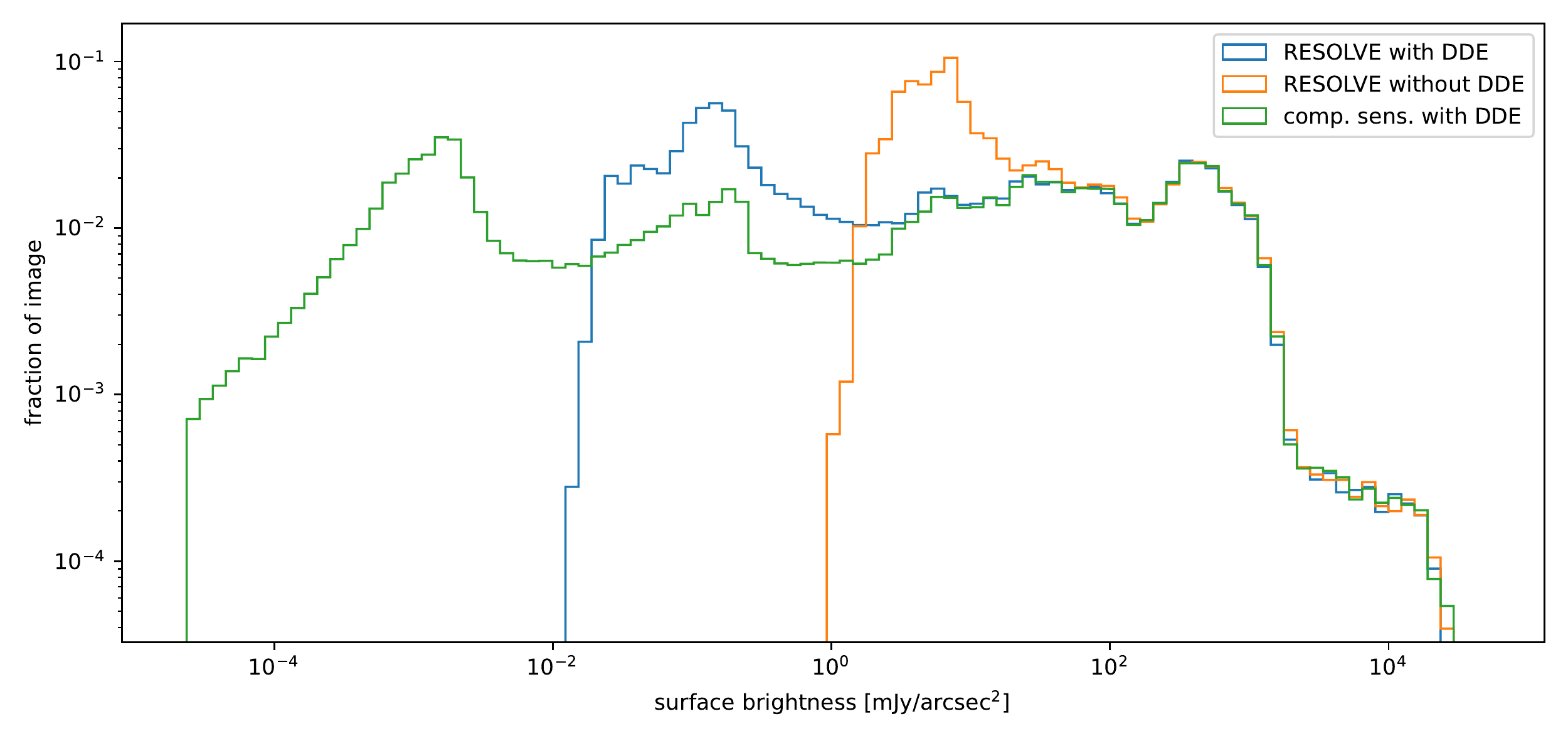}
        \caption{Histogram of surface brightness values of the \texttt{resolve} reconstruction with direction-dependent calibration, the \texttt{resolve} reconstruction of \cite{Arras2021} with classical calibration, and the compressed sensing reconstruction from \cite{Dabbech_2021} also with direction-dependent calibration. The histogram is computed from the images displayed in Fig. \ref{fig:cyg_comp} showing on the $y$-axis the fraction of the image with this surface brightness.}
        \label{fig:hist}
\end{figure*}

\begin{figure*}
   \centering
      \includegraphics[width=17cm]{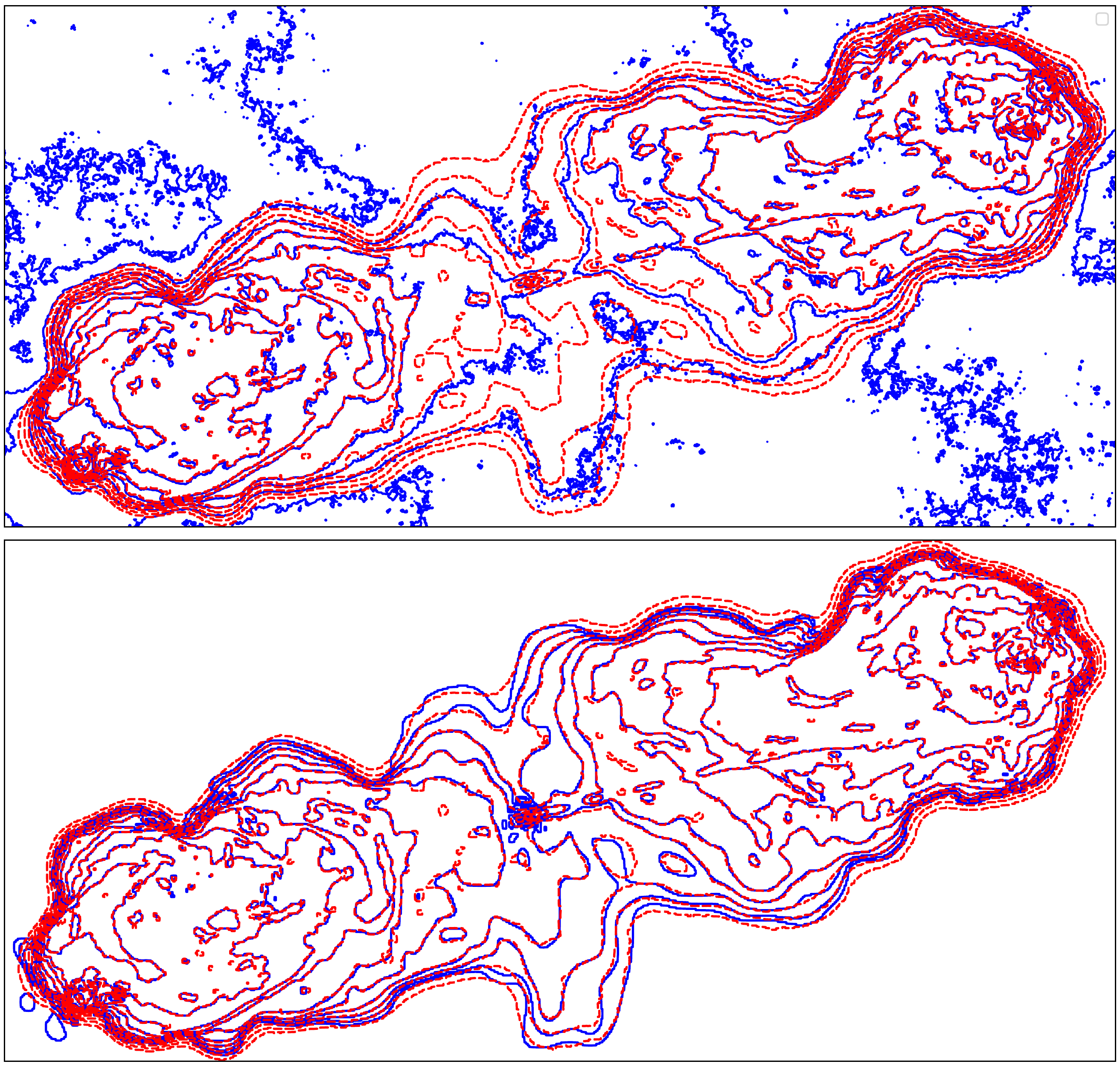}
        \caption{Flux contours of the Cygnus~A reconstructions. The surface brightness levels of the contours are logarithmically spaced and range from $1\, \text{mJy}/\text{arcsec}^{2}$ to $4.25\cdot 10^{4}\, \text{mJy}/\text{arcsec}^2$. Top panel: In solid blue \texttt{RESOLVE} reconstruction from \cite{Arras2021} without direction-dependent calibration. In dashed red \texttt{RESOLVE} reconstruction of this work with direction-dependent calibration. Bottom panel: In solid blue compressed sensing reconstruct of \cite{Dabbech_2021} with direction-dependent calibration. In dashed red again \texttt{RESOLVE} reconstruction of this work with direction-dependent calibration.}
        \label{fig:cyg_cont}
\end{figure*}

Besides the resolution, the sensitivity, specifying the dimmest recovered emission, is the second important quantity determining the quality of a radio map. This sets the dynamic range of a radio map as the ratio between the faintest to the brightness emission.
By comparison of three images in Fig. \ref{fig:cyg_comp}, it is evident that the two algorithms performing direction-dependent calibration provide radio maps with high dynamic range, as they recover structures with much lower surface brightness. The \texttt{resolve} reconstruction of \cite{Arras2021} without direction-dependent calibration has a significantly higher residual background brightness. The dynamic range gain also manifests in Fig. \ref{fig:hist}, which depicts as a histogram the fraction of the images with a given surface brightness. For a dynamic range of about $2.5$ orders of magnitude, all three reconstructions predict the same fraction for the given surface brightness. For surface brightnesses lower than $50\, \text{mJy}/\text{arcsec}^{2}$, only the two algorithms, including direction-dependent calibration, deliver consistent results. In contrast, the \texttt{resolve} reconstruction of \cite{Arras2021} utilizing classically calibrated data deviates significantly from the other two methods. The accurate agreement of the \texttt{resolve} reconstruction of this work and the compressed sensing reconstruction of \cite{Dabbech_2021} can also be confirmed by the contour plots in Fig. \ref{fig:cyg_cont}. The contour lines, spanning a surface brightness range of $4.3$ orders of magnitude, are in close agreement between the two reconstructions. In contrast, the flux contours of the \texttt{resolve} reconstruction from \cite{Arras2021} show significant deviations in the low surface brightness regions.

Thus, to conclude, ignoring direction-dependent effects in the calibration solutions limits the dynamic range of the recovered radio maps. The Bayesian \texttt{resolve} reconstructions utilizing direction-dependent calibration significantly improve the dynamic range, and the results are compatible with other work.

\begin{figure*}
   \centering
      \includegraphics[width=17cm]{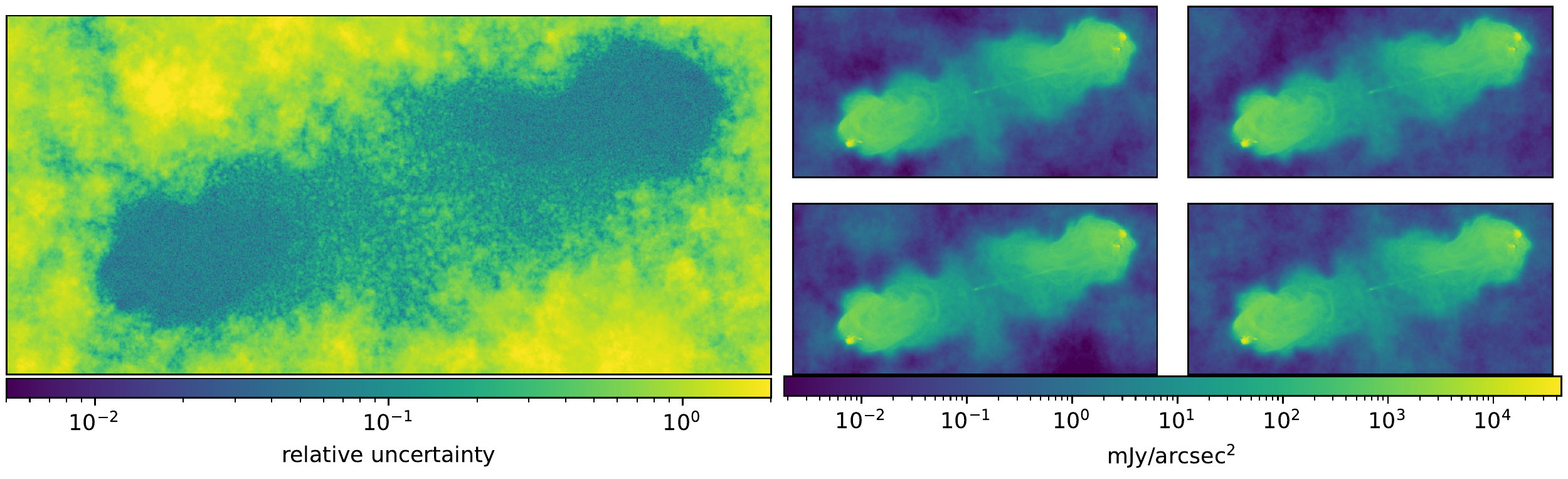}
        \caption{Left column: Relative uncertainty of the  reconstruct sky brightness computed from posterior samples. Two right columns: Four representative posterior samples of the geoVI reconstructions.}
        \label{fig:err}
\end{figure*}

The geoVI algorithm used in the \texttt{resolve} framework outputs a set of posterior samples of the sky brightness (see Sec. \ref{sec:geoVI}). The posterior mean sky brightness can be computed from these samples, as shown in the previous plots. Besides the mean, also other summary statistics of the sky brightness posterior distribution can be computed, such as the uncertainty. The left column of Fig. \ref{fig:err} depicts the relative uncertainty. In the regions of the hot spots of Cygnus~A, the relative uncertainty is on the order of a few percent. Towards the fainter ends of the lobes, the uncertainty rises towards $10\%$ or even higher in the faintest regions. Outside the contours of Cygnus~A, the relative uncertainty is significantly larger and exceeds $100\%$. This is expected since the surface brightness is negligibly small in these regions. Also, the \texttt{resolve} reconstruction of \cite{Arras2021} provides uncertainty estimates, which are compatible with the uncertainty estimates of this work. The compressed sensing method of \cite{Dabbech_2021} cannot estimate the uncertainty. 

The posterior distribution can also be analyzed based on individual samples. A set of four representative posterior samples is shown in Fig. \ref{fig:err} in the right column. While the samples show only tiny variations in high surface brightness regions, they are more diverse in regions with lower brightness, indicating higher posterior uncertainty.

\subsection{Calibration solutions}
In this section, we will briefly discuss the antenna gain solutions. As our antenna gain model (\ref{sec:model}) factorizes the antenna gains into a purely time-dependent term $g_p(t)$, an only direction-dependent $g_p(\boldsymbol{l})$, and a jointly direction and time-dependent term $g_p(\boldsymbol{l}, t)$, we can analyze the reconstruction of these terms separately. As it will turn out in the following sections, the statistical properties of the three factors are very different.
As we model each of the factors with separate Gaussian processes, inferring their power spectra, the algorithm can capture the different statistics and exploit them to further constrain the reconstructions.

\subsubsection{Only time-dependent gains: $g_p(t)$}
\begin{figure}
   \resizebox{\hsize}{!}{\includegraphics{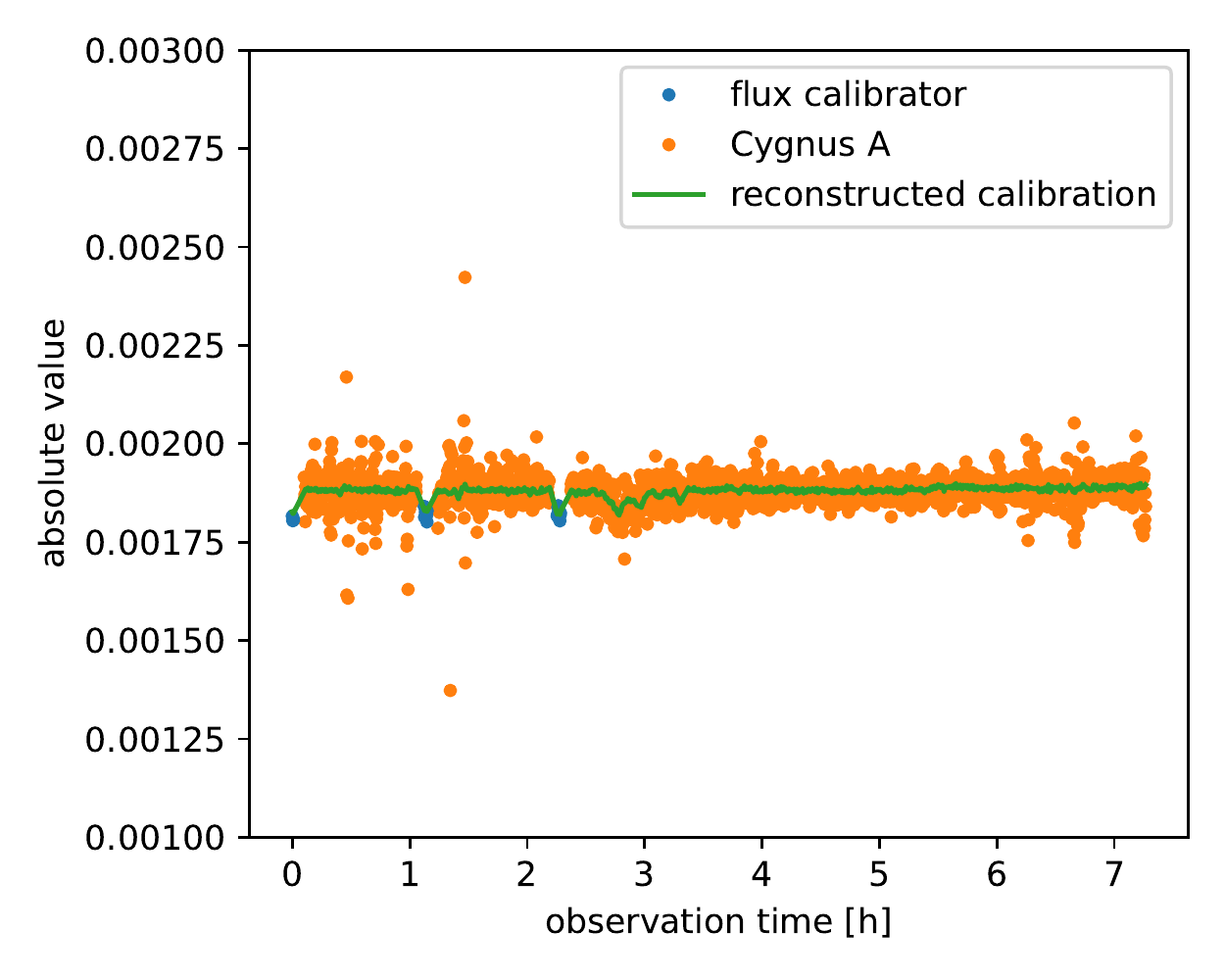}}
   \caption{Green: Absolute value of the reconstructed time-dependent, but direction-independent, calibration for one baseline. The blue and orange dots are the absolute values of the visibilities of the flux calibrator and target observations, projected to the calibration space by dividing out their respective sky models. More specifically, blue dots: Flux calibrator visibilities divided by the model visibilities computed from the flux calibrator sky model without time-dependent calibration factors. Orange dots: Cygnus~A visibilities divided by the model visibilities computed from the Cygnus~A sky model, including the direction-dependent calibration but excluding the purely time-dependent calibration. The offset between the calibration data points and the target data points makes it necessary to allow for possible discontinuities in the antenna gain reconstructions when switching the observation target. Partly this offset might also be due to a oversimplified flux calibrator model.}
   \label{fig:jumps}
\end{figure}

In Fig. \ref{fig:jumps}, the absolute value of the only time-dependent part of the calibration solution is depicted for one baseline, thus in formulas $\text{abs}(g_{p}(t)g^{*}_{q}(t))$. The absolute value of the antenna gain, thus the flux scale, is relatively stable over time. To visualize the jumps in the calibration solutions discussed in the previous section, we display the absolute value of the visibilities of the flux calibrator as well as the science target Cygnus~A projected to the calibration space. More precisely, we divide the visibilities of the respective observation by the model visibilities computed from the sky model without the only time-dependent calibration factor $g_{p}(t)g^{*}_{q}(t)$. Thus, as a formula:
\begin{align}
   V_{pqt}^{\text{projected}} = \frac{V_{pqt}}{\int C(\boldsymbol{l},w_{pqt})I(\boldsymbol{l}) \widetilde{G}_{pt}(\boldmath{l}, t) \widetilde{G}^{*}_{qt}(\boldmath{l}, t) e^{-2\pi i (\boldsymbol{k}_{pqt}\cdot\boldsymbol{l})} d\boldsymbol{l}}
\end{align}
with $V_{pqt}$ being the visibilities of the observation, $C(\boldsymbol{l})$ the $w$-term correction factor, $I(\boldsymbol{l})$ the reconstructed sky model and 
\begin{align}
   \widetilde{G}_{pt}(\boldmath{l}, t) \widetilde{G}^{*}_{qt}(\boldmath{l}, t)=g_{p}(\boldsymbol{l},t)g^{*}_{q}(\boldsymbol{l},t) g_{p}(\boldsymbol{l})g^{*}_{q}(\boldsymbol{l})
\end{align}
the calibration solution excluding the only time-dependent part $g_{p}(t)g^{*}_{q}(t)$. The offset between the visibilities from the flux calibrator and the visibilities of the Cygnus~A observation is clearly visible in Fig. \ref{fig:jumps}, and is on the order of a few percent. The discontinuities in the calibration solution, as described in Sec. \ref{sec:model}, were introduced to model these offsets.

\subsubsection{Only direction-dependent gains: $g_p(\boldsymbol{l})$}
Fig. \ref{fig:dde_ct} depicts the direction-dependent but time independent gain $g^{ll/rr}_{p}(\boldsymbol{l})$ of one antenna. The phase and amplitude variations are both on the level of a few percent. The interferometer is only sensitive to variations of the gain solutions in directions with significant sky surface brightness since the antenna gains are a multiplicative effect on the sky. Therefore, to indicate which regions of the reconstructed gains solutions actually affect the likelihood, we superimpose the flux contours of Cygnus~A on the plot of the calibration solution.

\subsubsection{Jointly time and direction-dependent gains: $g_p(\boldsymbol{l}, t)$}
\begin{figure}
   \resizebox{\hsize}{!}{\includegraphics{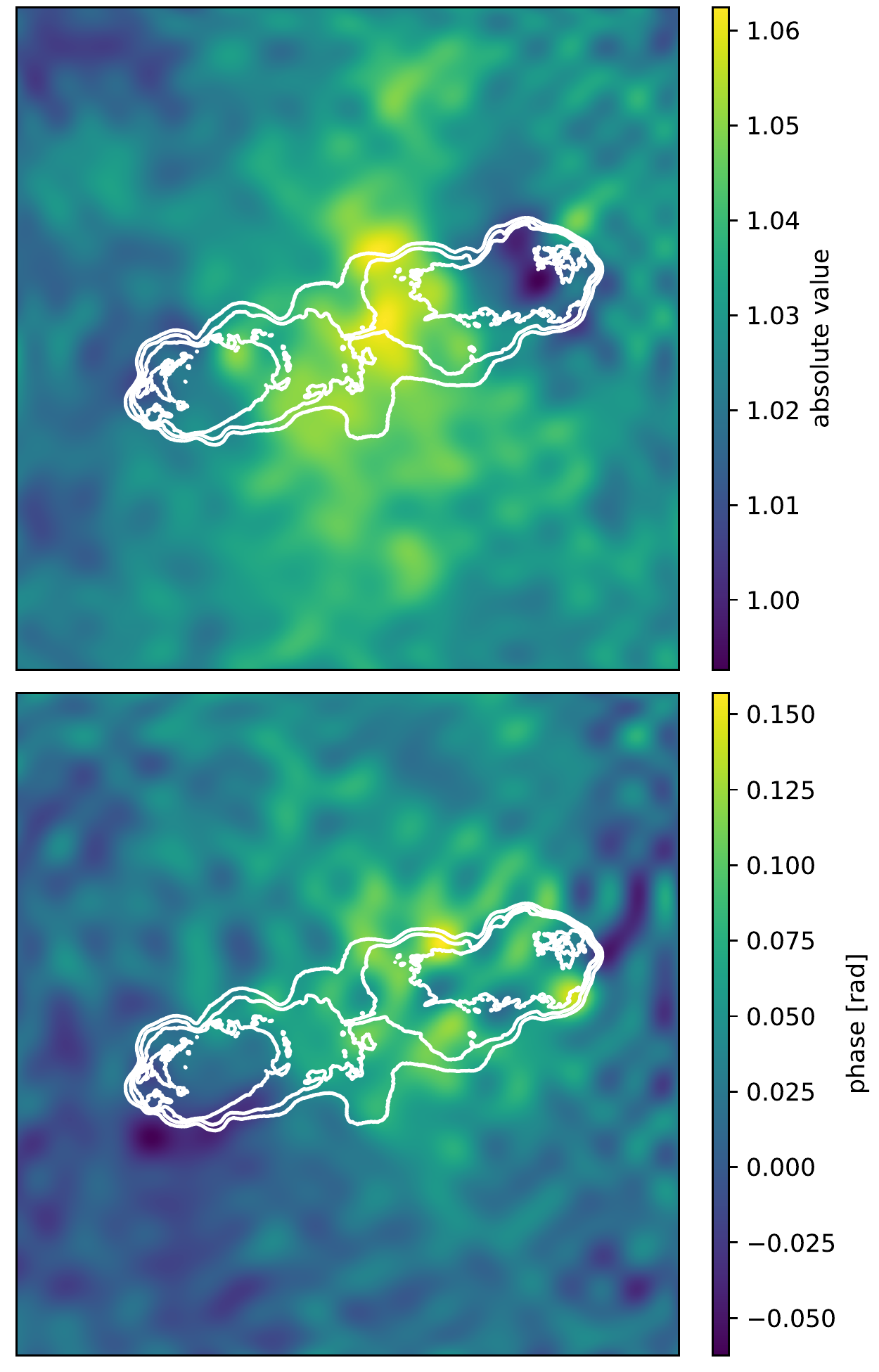}}
   \caption{Phase and amplitude of the direction-dependent, time-independent gain factor for one antenna. On top the flux contours of Cygnus~A are indicated in white, since the interferometer is only sensitive to direction-dependent effects in regions with significant surface brightness. Thus, gain values outside the contours are neither accurate nor relevant.}
   \label{fig:dde_ct}
\end{figure}

\begin{figure*}
   \centering
      \includegraphics[width=17cm]{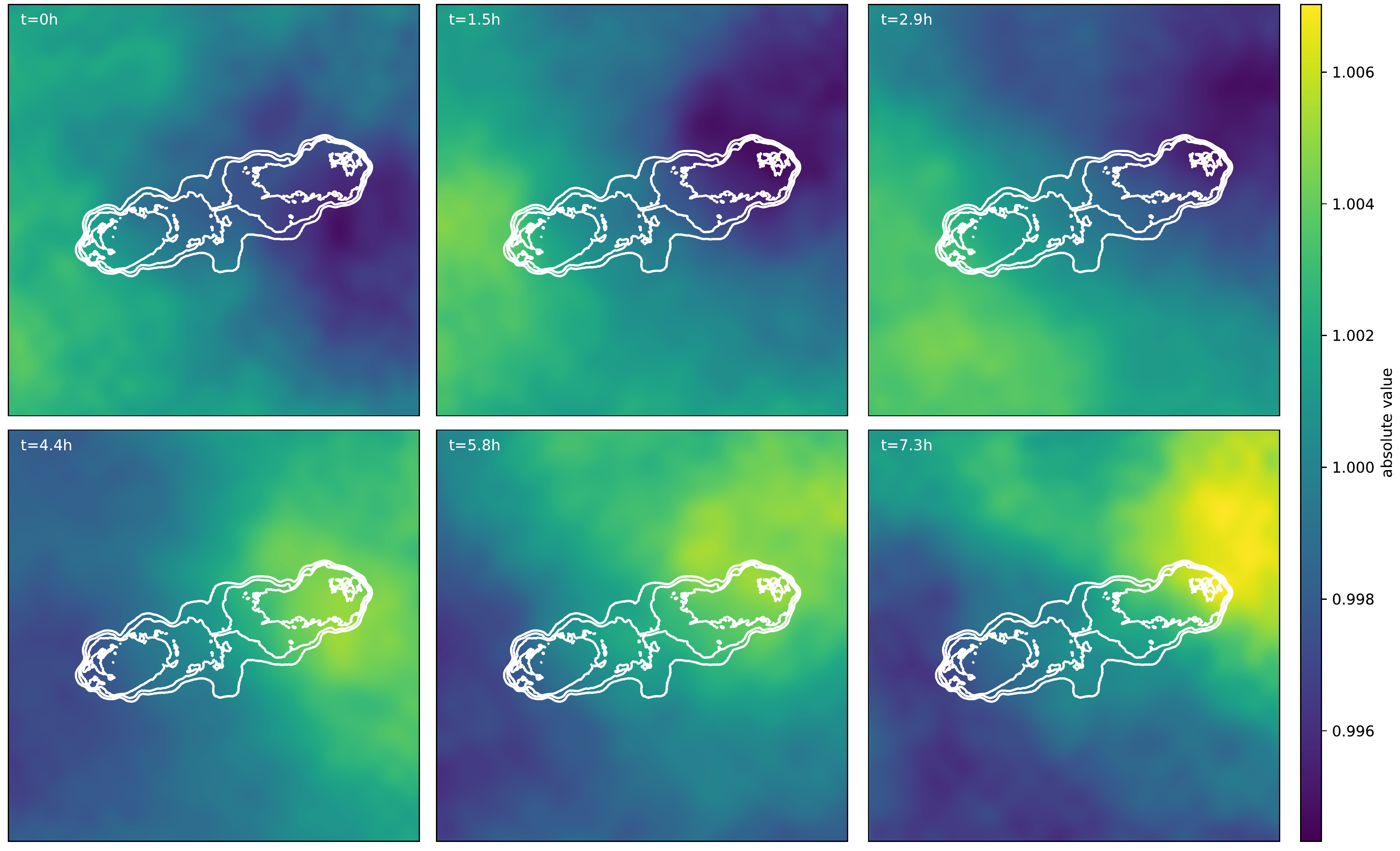}
        \caption{Absolute value of jointly time and direction-dependent gain of antenna 0, depicted at six timesteps of the observation. On top, the contours of Cygnus~A are shown since the interferometer is only sensitive to the direction-dependent gain variations in regions with significant sky brightness. Gain values outside the contours are neither accurate nor relevant.}
        \label{fig:dde_joint}
\end{figure*}

The absolute value of the jointly time and direction-dependent antenna gains $g^{ll/rr}_{p}(\boldsymbol{l}, t)$ from which the only time and only direction-dependent variations are excluded is displayed in Fig. \ref{fig:dde_joint}. Similar to Fig. \ref{fig:dde_ct}, the flux contours of Cygnus~A are superimposed to indicate which regions actually have a significant effect. The variations of the jointly direction and time-dependent gains are less than a percent and therefore smaller than the only time and only direction dependent effects. The spatial correlation length is, compared to the only direction-dependent gain solutions, significantly larger. This result underlines that a split of the antenna gains into an only direction and a jointly direction and time-dependent part can be beneficial because the generative model for $g^{ll/rr}_{p}(\boldsymbol{l}, t)$ can exploit this longer correlation length.

A direct quantitative comparison with the work of \cite{Dabbech_2021}, as done for the Cygnus~A sky images, is not possible for the calibration solutions, as the calibration model in \cite{Dabbech_2021} is different and the results are only published in the form of images. Nevertheless, the scale of the amplitude and phase variations are comparable. Also, the correlation length of the jointly direction and time-dependent gain solutions is similar to the correlation length of the direction and time-dependent calibration solutions in \cite{Dabbech_2021}. However, in the work of \cite{Dabbech_2021}, the spatial and temporal correlation lengths are set somewhat externally by specifying the support size of the corresponding Fourier kernels. In contrast, in this work, the correlation structures are not fixed but reconstructed from the data.

\begin{figure*}
   \centering
      \includegraphics[width=17cm]{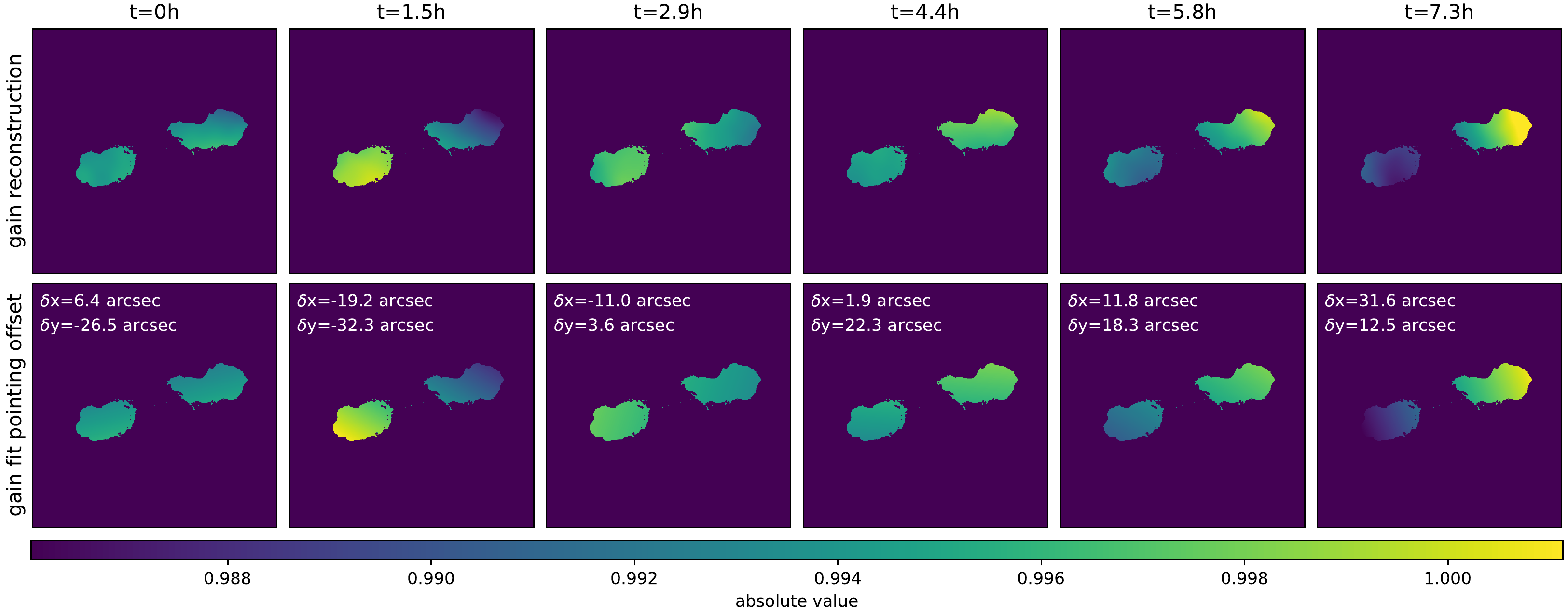}
        \caption{Top row: Absolute value of jointly time and direction-dependent gain of antenna 0 (same as Fig: \ref{fig:dde_joint}) restricted to sky region with significant flux. Bottom: Fit of the antenna gain resulting from pointing errors to the reconstructed gain in the top row. Except for minor differences, the gains resulting from pointing errors can reproduce the reconstructed gains. The fitted offset to the nominal pointing direction is specified as $\delta x$ and $\delta y$. As Cygnus A is less extended in $y$-direction than in $x$-direction, the gains are much less sensitive to pointing offsets in $y$-direction.}
        \label{fig:dde_pointing}
\end{figure*}

Pointing errors are expected to be an important source of jointly time and direction-dependent gains. The antenna beam is falling off with increasing distance to the center of the field of view. This makes the sky to appear brighter in the direction of the pointing offset and dimmer in the opposite direction, leading to a direction-dependent antenna gain. Mathematically the direction-dependent antenna gain can be computed by
\begin{align}\label{eq:pointing}
   g_{p}(\boldmath{l},t) = \frac{B_{p}(\delta x_t, \delta y_t)}{B_{p}(0, 0)},
\end{align}
with $B_{p}(\delta x_t)(\delta y_t)$ being the beam of antenna $p$ with an pointing offset of $(\delta x_t, \delta y_t)$, and $B_{p}(0, 0)$ the beam of the nominal pointing. Using Eq. \ref{eq:pointing}, we fitted pointing offsets to the absolute value of the direction-dependent gains, resulting in pointing offsets of up to $30\, \text{arcsec}$ as depicted in Fig. \ref{fig:dde_pointing}. As visible in Fig. \ref{fig:dde_pointing}, the gains resulting from pointing offsets relatively closely follow the reconstructed gain. Nevertheless, some deviations remain as pointing errors are not the only source for time and direction-dependent gains.

\section{Conclusions}\label{sec:conclusion}
This paper introduces a novel Bayesian radio interferometric imaging and calibration algorithm based on the \texttt{resolve} framework. Calibration and imaging are performed jointly. The calibration includes corrections for direction-dependent effects. To this goal, the antenna gain solutions are split into three factors, a purely time-dependent term, a purely direction-dependent term, and a jointly direction and time-dependent term. Having a separate model for each factor of the gain solutions allows to exploit the different statistics of the individual factors.

We demonstrate the method on VLA's Cygnus~A observation, a widely used target for benchmarking novel algorithms. In comparison with previous work \citep{Arras2021}, imaging Cygnus~A also with the \texttt{resolve} framework but classically calibrated data, the dynamic range is significantly increased. At the same time, in high and medium surface brightness regions, the two reconstructions are closely consistent. When comparing with \cite{Dabbech_2021}, a compressed sensing method including direction-dependent calibration, the results not only agree in high and medium surface brightness regions but also in areas with very low flux. As the reconstruction algorithms of the sky radio maps of this work and \cite{Dabbech_2021} are based on very different methodologies, namely Bayesian inference and compressed sensing, their close agreement is a strong indication of their reliability even at low surface brightness regions. Furthermore, this underlines the necessity of an accurate calibration model for obtaining high-fidelity radio maps with a large dynamic range. An accurate calibration model will become a critical component for exploiting the full potential of upcoming radio interferometers.

\begin{acknowledgements}
J.R. and P.A. acknowledge financial support by the German Federal Ministry of Education and Research (BMBF) under grant 05A20W01 (Verbundprojekt D-MeerKAT).
. \end{acknowledgements}

\bibliographystyle{aa}
\bibliography{main}

\end{document}